\documentclass[aps,prl,twocolumn,floatfix,showpacs,preprintnumbers,amsmath,amssymb,10pt]{revtex4-1}

\usepackage[dvips]{graphics}

\DeclareMathAlphabet{\mathsfit}{\encodingdefault}{\sfdefault}{m}{sl}
\SetMathAlphabet{\mathsfit}{bold}{\encodingdefault}{\sfdefault}{bx}{sl}

\usepackage[dvips]{graphics}
\usepackage{geometry}
 \usepackage{ctable}
 \usepackage[percent]{overpic}
\usepackage{textgreek}
\usepackage{bm}
\usepackage{contour}
\usepackage{braket}
\usepackage{soul}

\usepackage{blindtext}
\usepackage[percent]{overpic}
\usepackage{mathtools}
\usepackage{xcolor}
\usepackage{multirow}
\usepackage{array}

\usepackage[percent]{overpic}

\usepackage{mathtools}
\renewcommand{\vec}[1]{\mathbf{#1}}

\renewcommand{\vec}[1]{\mathbf{#1}}

\newcommand{\tens}[1]{\mbox{\textsf{\textbf{#1}}}}

\newcommand{\Greektens}[1]{\contour[3]{black}{#1}}

\newcommand{\dif}{\!\! \mathrm{d}}
\newcommand{\mi}{\textrm{i}} 
\newcommand{\me}{\mathrm{e}}

\newcommand{\Epw}{\vec{E}_\textrm{p}(\omega)}
\newcommand{\Eprw}{\vec{E}_\textrm{p}(\vec{r},\omega)}

\newcommand{\beamWaist}{\textrm{w}}
\newcommand{\beamwaist}{\textrm{w}}
 \newcommand{\kPerp}{k_\perp}

\renewcommand{\vr}{\vec{r}}
\newcommand{\rp}{\vec{r}^\prime}

\newcommand{\intzi}{\int_0^\infty}

\begin{document}

\title{Theory of quantum-vacuum detection}

\author{Frieder Lindel$^1$}
\author{Robert Bennett$^{1}$}
\author{Stefan Yoshi Buhmann$^{1}$}
\affiliation{$^1$ Physikalisches Institut, Albert-Ludwigs-Universit\"at Freiburg, Hermann-Herder-Stra{\ss}e 3, 79104 Freiburg, Germany}

\date{\today}

%\begin{abstract} 
%Yet to come... 
%
%\end{abstract}

\begin{abstract}
Recent progress in electro-optic sampling has allowed direct access to the fluctuations of the electromagnetic ground state. Here, we present a theoretical formalism that allows for an in-depth characterisation and interpretation of such quantum-vacuum detection experiments by relating their output statistics to the quantum statistics of the electromagnetic vacuum probed. In particular, we include the effects of absorption, dispersion and reflections from general environments. Our results agree with available experimental data while leading to significant corrections to previous theoretical predictions and generalises them to new parameter regimes. Our formalism opens the door for a detailed experimental analysis of the different characteristics of the polaritonic ground state, e.g. we show that transverse (free-field) as well as longitudinal (matter or near-field) fluctuations can be accessed individually by tuning the experimental parameters. 
\end{abstract}

\maketitle

Over ninety years ago, Heisenberg formulated the uncertainty principle \cite{Heisenberg1927}. One of its most fascinating consequences appears in quantum electrodynamics (QED), where the commutation relations imply so-called zero-point fluctuations of the electromagnetic field, persisting even in the ground state of the theory: the quantum vacuum. It has been argued that indirect evidence for these fluctuating fields can be seen in experiments measuring spontaneous decay rates \cite{Drexhage1970}, the Lamb shift \cite{Lamb1947} or the Casimir force \cite{Casimir1948}. These effects are not only of fundamental interest in the context of studying the vacuum field, but also play an important role in many different areas of science, such as nanotechnology \cite{Serry1998} and adhesion \cite{Brivio1999}. % In all these scenarios, the quantum ground state of the electromagnetic field is altered by the presence of media leading to the medium-assisted ground state of QED consisting of matter and free field fluctuations [][]. %%Nevertheless, as argued in Refs.~\cite{Schwinger1975, Milonni1982,Milonni1992,Milonni1994,Jaffe2005}, all of the above mentioned effects are not necessarily definitive proof of the existence of the fluctuating vacuum field, let alone its change induced by surfaces. An alternative mechanism offered in those works is radiation reaction, where matter interacts with the field induced by its own fluctuating dipoles. \\
Recently, experiments based on nonlinear optics have opened up an alternative route to the ground state of the electromagnetic field \cite{Riek2015,Benea-Chelmus2019}. In nonlinear optics photons can effectively be made to interact with each other \cite{Boyd2003,Scheel2006,Chang2014a} which has become an integral component of a wide range of experimental techniques \cite{Franken1961,Mukamel1995,Kwiat1995} and permits remarkable insights into fundamental physics \cite{Anderson1995a,Silberhorn2001}. These new experimental techniques include electro-optic sampling \cite{Riek2015,Benea-Chelmus2019} with a non-linear crystal or the use of a time-dependent refractive index (the dynamical Casimir effect) \cite{Westerberg2018,Vezzoli2018}. 

\begin{figure}
\centering
\includegraphics[width=0.8\columnwidth]{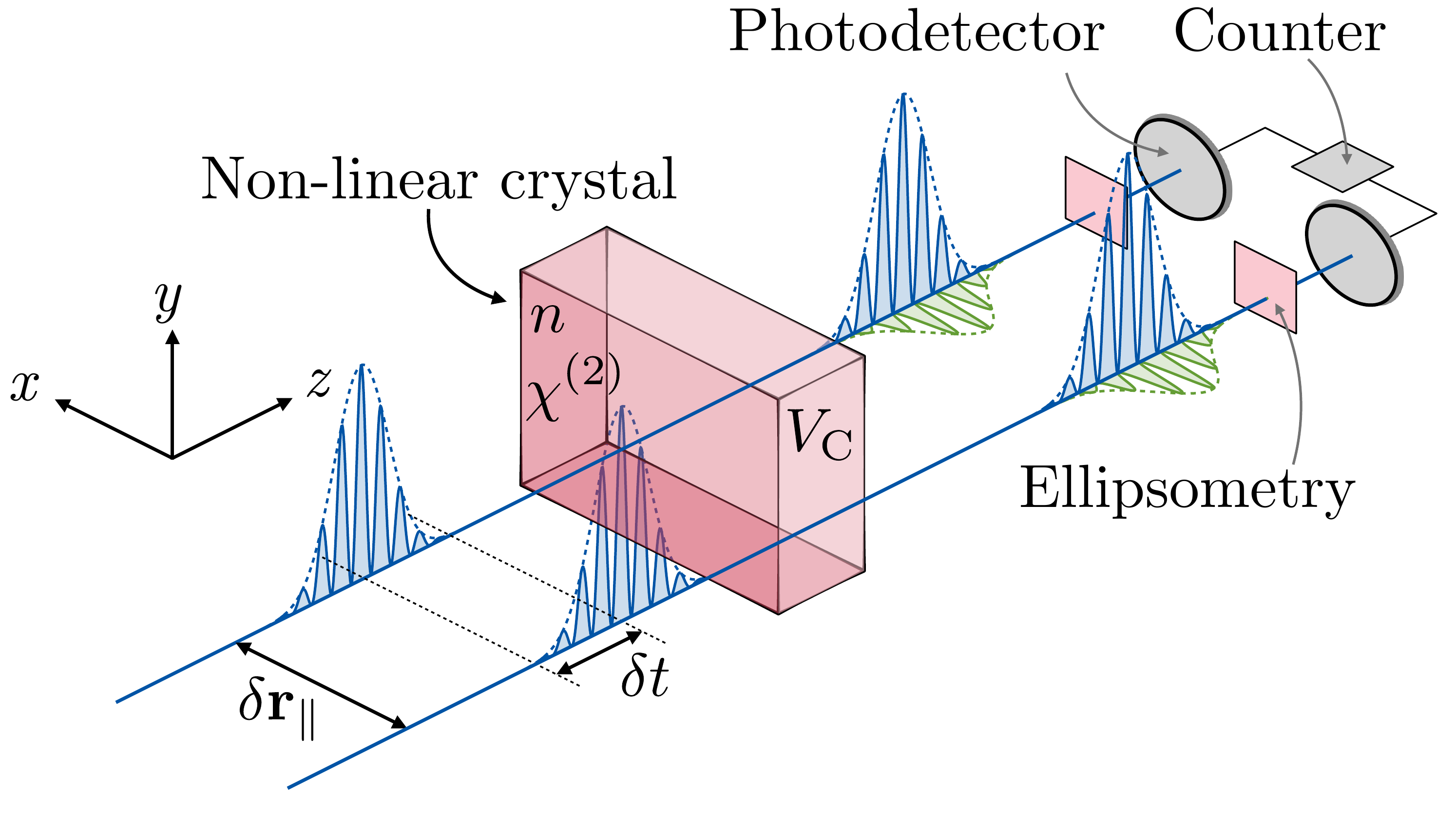} 
\caption{\textit{Experimental setup of correlation measurement of the quantum vacuum via electro-optic sampling:} Two linearly polarised laser pulses with mutual offset $\delta \vec{r}_\parallel$ and delay $\delta t$ propagate through a nonlinaer crystal with refractive index $n$, nonlinear susceptibility $\chi^{(2)}$ and volume $V_C$. The pulses mix inside the crystal via the nonlinear coupling with quantum fluctuations of the electromagnetic field, leading to a polarisation shift of the laser pulses. Via an ellipsometry analysis of the pulses emerging from the crystal, it is possible to observe the quantum vacuum and to reveal the spectrum of the two-point correlation function of the electric-field operator in the polaritonic ground state inside the crystal \cite{Benea-Chelmus2019}. }
\label{fig:GeneralDiagram}
\end{figure}

In electro-optical sampling, a linearly polarised, ultra-short laser pulse propagates through an non-linear crystal which mixes the laser pulse with any ambient electric field via its nonlinear properties \cite{Boyd2003} to form a new electric field. This leads to a change of the pulse's polarisation so that by analysing the polarisation of the field emerging from the crystal one obtains information about the ambient field inside it \cite{Valdmanis1986,Riek2017}. The sensitivity of this setup to extremely weak electric fields allows one to measure the effect of the fluctuating vacuum upon the output statistics \cite{Riek2015,Moskalenko2015}, providing direct access to zero-point fluctuations. Using two such laser pulses (see Fig.~\ref{fig:GeneralDiagram}), it is possible to retrieve information about correlations of the QED vacuum between distinct spatio-temporal regions and this way access its spectral decomposition \cite{Benea-Chelmus2019}, for example. 
Following the pioneering works using such setups \cite{Riek2015,Riek2017,riek2017subcycle,Benea-Chelmus2019} and the accompanying theoretical analyses \cite{Moskalenko2015,guedes2019spectra,kizmann2019subcycle,moskalenko2019correlations} the question regarding the nature of the quantum fluctuations accessed has been raised \cite{ScienceComment}. In particular, as electro-optic sampling is necessarily carried out inside a nonlinear optical crystal, the relation of the sampled quantum vacuum to the paradigmatic free-space vacuum is an important question. Here, we address this question and offer 
%  \\
%Here, we remove these limitations in 
a general theoretical framework based on macroscopic QED
%, which can be used as a starting point in for a detailed analysis of the quantum vacuum 
which provides a basis for a detailed characterisation and intepretation of quantum-vacuum detection
via electro-optic sampling. Our theory is capable of predicting the output statistics of such experiments, accounting for inhomogeneous dispersive and absorptive media by considering the full medium-assisted ground state of the system as predicted by linear QED consisting of composite (polariton-like) matter and free-field fluctuations --- the vacuum which is probed is the \textit{polaritonic} vacuum
%SYB: added
which generalises the free-space vacuum to account for the nonlinear-crystal environment.
We further show that this formalism agrees well with experimental data while introducing significant advances over previous theoretical frameworks. Our formalism also provides new fundamental insights---for example we show that by tuning the parameters of the experimental setup within a realistic range, one can individually address longitudinal (matter-like) and transverse (free field-like) ground state fluctuations. \\

We begin with a brief account of the underpinnings of our theory, presented in detail in Ref.~\cite{PRA}. The propagation of a coherent laser pulse through a medium with second-order nonlinearity induces a nonlinear polarisation field given by \cite{Boyd2003} \mbox{$ \hat{\vec{P}}_\textrm{NL}(\vec{r},\omega)\!=\!\!\! \int_{-\infty}^\infty \! \dif \Omega \,\, \! \Greektens{$\chi$}^{(2)}\!(\vec{r}, \Omega, \omega\! -\! \Omega)  \! \star \!   \hat{\vec{E}}(\vec{r},\Omega) \hat{\vec{E}}(\vec{r},\omega \! -\! \Omega ).$}
%\begin{align} \label{eq:PNLinear}
%\hat{\vec{P}}_\textrm{NL}(\vec{r},\omega)\!=\!\!\!\!\! \int\limits_{-\infty}^\infty \! \dif \Omega \,\, \! \Greektens{$\chi$}^{(2)}\!(\vec{r}, \Omega, \omega\! -\! \Omega)  \! \star \!   \hat{\vec{E}}(\vec{r},\Omega) \hat{\vec{E}}(\vec{r},\omega \! -\! \Omega ).
%\end{align} 
Here, $\Greektens{$\chi$}^{(2)}$ is the nonlinear susceptibility tensor of the medium and we have defined a shorthand \mbox{$(\Greektens{$\chi$}^{(2)} \! \star \! \hat{\vec{E}}  \hat{\vec{E}})_i \equiv \sum_{jk} \chi^{(2)}_{ijk} \hat{E}_j \hat{E}_k $}. The nonlinear polarisation acts as an additional source term in the wave equation for the electric field, which can be formally solved as a Lippmann-Schwinger equation
\begin{multline} \label{eq:LippmannSchwinger}
\hat{\vec{E}}(\vec{r},\omega) = \hat{\vec{E}}_\textrm{vac}(\vec{r},\omega) + \vec{E}_\textrm{p}(\vec{r},\omega)\\
 + \mu_0 \omega^2 \int\limits_{V_\textrm{C}} \dif^3 r^\prime \, \tens{G}(\vec{r},\vec{r}^\prime,\omega) \cdot  \hat{\vec{P}}_\textrm{NL}(\vec{r}^\prime,\omega) 
\end{multline}
where $\mu_0$ is the vacuum permeability and $\tens{G}(\vec{r},\vec{r}^\prime,\omega)$ is the Green's tensor of the vector Helmholtz equation \cite{SUPP} and $V_\text{C}$ is the volume of the non-linear crystal. Furthermore, we are working in the vacuum picture in which the coherent laser pulse is given by the sum of the vacuum field operator $\hat{\vec{E}}_\textrm{vac}(\vec{r},\omega)$ and a classical laser pulse $ \vec{E}_\textrm{p}(\vec{r},\omega)$ \cite{Knight1983,PRA}. Note that $ \vec{E}_\textrm{p}(\vec{r},\omega)$ may represent two spatially and temporally separated laser pulses, such as those featuring in the recent experiment \cite{Benea-Chelmus2019}. 

Equation %\st{\eqref{eq:PNLinear} and}
\eqref{eq:LippmannSchwinger} defines a formal solution for $\hat{\vec{E}}(\vec{r},\omega)$, but is infinitely recursive. To solve it we use a Born series, which can be seen as a perturbation expansion in $\Greektens{$\chi$}^{(2)}$ to the desired order, for details see Ref.~\cite{PRA}. From this procedure one obtains the electric field emerging from the nonlinear crystal as a function of the input fields and the Green's tensor describing its geometry and material response.
%This theoretical framework can be applied to various experimental setups requiring a fully quantised treatment of nonlinear optical processes. 

Here, we use our solution to Eq.~\eqref{eq:LippmannSchwinger} to find the output statistics of an electro-optic sampling experiment (see Fig.~\ref{fig:GeneralDiagram}). These are found from the variance of the electro-optical operator $\hat{S}$, which for the single-beam setup used in Ref.~\cite{Riek2015} reads \cite{Moskalenko2015};
\begin{multline} \label{eq:S2General}
\hat{S} =\!\!  \int_0^\infty \!\!\!\!\! \dif \omega \, A(\omega) \!\! 
\int \dif^2 r_\parallel \!\!  \left[ \mi \hat{E}_y^\dagger(\vec{r}_\parallel, \omega)  \hat{E}_x(\vec{r}_\parallel, \omega) \!  + \!  \text{h.c.}  \right],
\end{multline}
where $A(\omega) = 4\pi \epsilon_0  c n(\omega)\eta(\omega)/\hbar \omega $ and $\eta$ is the efficiency of the photodetector. For the more general setup used in Ref.~\cite{Benea-Chelmus2019} where two laser pulses $\vec{E}_{1,2}$ are used and which is also depicted in Fig.~\ref{fig:GeneralDiagram} one accesses the quantity \mbox{$\hat{S}^2(\delta t, \delta \vec{r}_\parallel) = (\hat{S}_1 \hat{S}_2 + \hat{S}_2\hat{S}_1)/2$}. Here, $\hat{S}_i$ is defined as in Eq.~\eqref{eq:S2General} with the replacement  $\vec{E}_\textrm{p} \to \vec{E}_i$. Note that $\hat{S}^2(0,0)  = \hat{S}^2$.
Using the perturbation expansion outlined above up to second order in $\Greektens{$\chi$}^{(2)}$ we can evaluate $\hat{S}^2(\delta t, \delta \vec{r}_\parallel)$ and find \cite{PRA}
\begin{multline}\label{eq33:Scor2OfFilter}
\langle : \hat{S}^2(\delta t, \delta \vec{r}_\parallel) : \rangle = \intzi \!\!\!\dif \Omega \intzi \!\!\! \dif \Omega^\prime \int_{V_\textrm{C}}\!\!\!\dif^3 r^\prime \int_{V_\textrm{C}} \!\!\! \dif^3 r^{\prime\prime} \\
\times \langle E_{\textrm{vac},x}(\vec{r}^\prime, \Omega)E^\dagger_{\textrm{vac},x}(\vec{r}^{\prime\prime}, \Omega^\prime) \rangle  F(\vec{r}^\prime, \vec{r}^{\prime\prime},\Omega,\Omega^\prime),
\end{multline}
with the field correlation function given through macroscopic QED as \cite{Scheel_mac_2009,Buhmann2012a}
\begin{multline} \label{eq:VacCorrelations}
 \langle \hat{E}_{\textrm{vac},x}(\vec{r}^\prime,\Omega) \hat{E}_{\textrm{vac},x}^{\dagger}(\vec{r}^{\prime\prime},\Omega^\prime) \rangle = \frac{2\hbar \mu_0}{\pi}\Omega^2 \delta(\Omega - \Omega^\prime) \\
 \times \left[ \frac{1}{2} + n(\Omega, T)\right] \text{Im}\left[G_{xx}(\vec{r}^\prime, \vec{r}^{\prime\prime},\Omega) \right] .
\end{multline}
The filter function $F$ can be found in the supplementary materials \cite{SUPP}, and depends on the spatio-temporal probe beam profile, the relative spatial offset $\delta \vec{r}_\parallel$ and temporal delay $\delta t$, the optical and geometric properties of the crystal and its environment through the Green's tensor and the linear part of the crystal's permittivity, accounting for dispersion and absorption. It determines which spatial and spectral parts of the vacuum field are accessed via this quantum-vacuum detector, see Fig.~\ref{fig:Comparison}. The simplest example of this filter function is that for a single laser pulse with a Gaussian profile and beam waist $\textrm{w}$, taken at equal frequencies and and neglecting absorption \cite{PRA};
\begin{multline}\label{eq:FilterPPGRI}
F(\vec{r}^\prime,\vec{r}^{\prime\prime}, \Omega,\Omega) =  \left( \frac{2 |\chi^{(2)}| c \mu_0 N \omega_p }{\beamWaist^2 n}\right)^2  f(\Omega)^2 \\
\times\me^{-(\vec{r}_\parallel^{\prime 2} +\vec{r}_\parallel^{\prime\prime 2})/\beamWaist^2} \me^{-\mi n_g \frac{\Omega}{c}(z^\prime-z^{\prime\prime})} .
\end{multline}
Here $N$ is the total number of detected photons, $\omega_p$ is the average detected frequency, $n$ the refractive index at the central frequency of the pulse $\omega_\textrm{c}$, $n_g$ the group refractive index and $f(\Omega)$ is the spectral autocorrelation function \cite{Moskalenko2015,SUPP}. \\
\begin{figure}
\centering
\includegraphics[width=0.8\columnwidth]{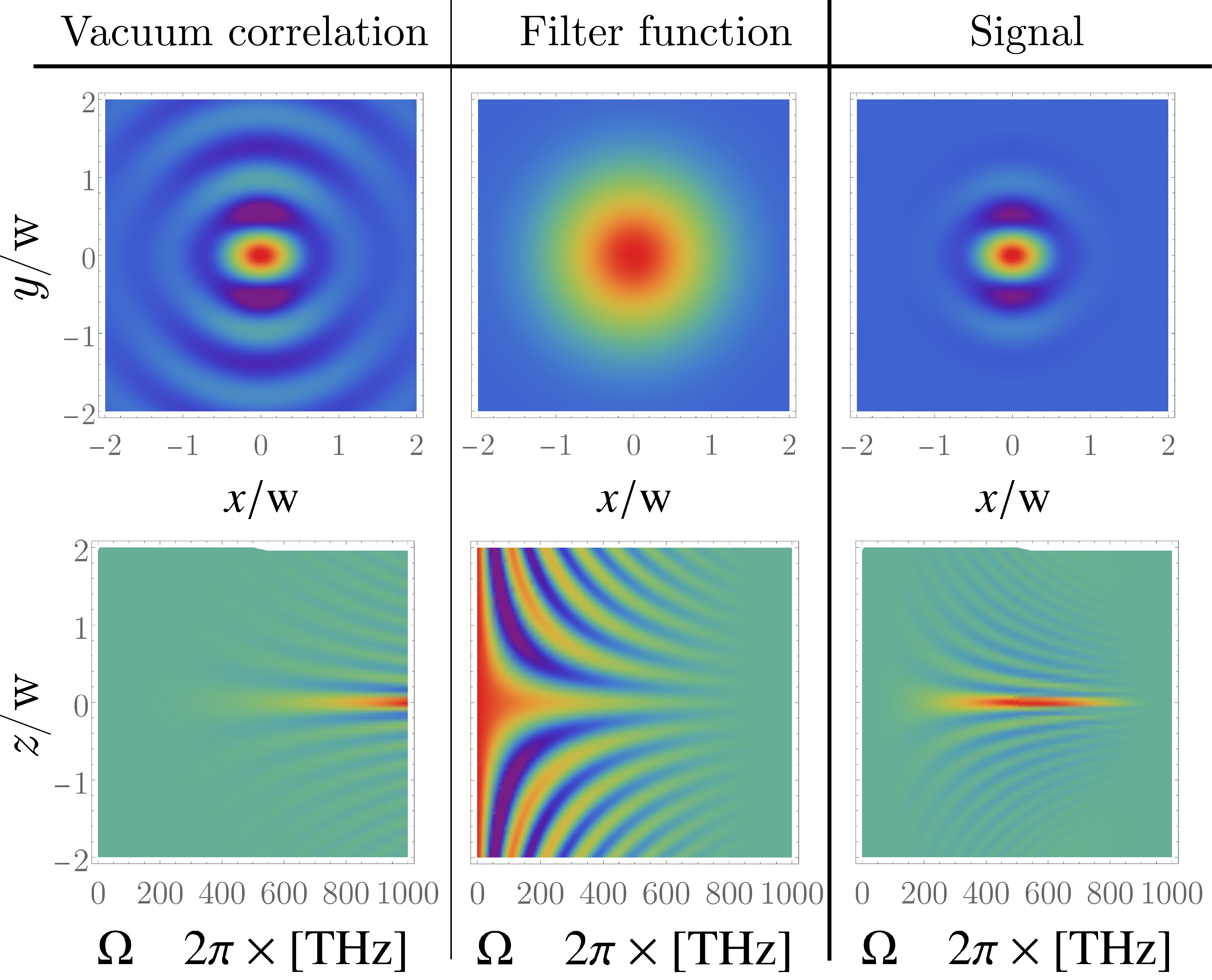} 
%  \begin{tabular}{@{} cc @{}}
%    \toprule
%    Approximation & $\langle :S^2: \rangle/\langle :S^2: \rangle_{\textrm{full}}$  \\ 
%    \midrule
%    Paraxial & 1.52 \\ 
%    Paraxial with cutoff & 1.26 \\ 
%    Taylor & 1.06 \\ 
%    Laser Paraxial & 1.03 \\ 
%    \bottomrule
%  \end{tabular}
\caption{\textit{Distinguishing contributions from the vacuum and from the filter function:} We consider one $y$ polarised laser pulse with Gaussian beam profile with beam waist $\beamwaist = 3\,\mu\textrm{m}$ and duration $\Delta t = 5.9\,\textrm{fs} $ to propagate through a ZnTe crystal with length $L = 7\mu$m \cite{SUPP}. Using Eqs.~\eqref{eq33:Scor2OfFilter}--\eqref{eq:FilterPPGRI} we plot the normalised filter function $F(\vr,\rp,\Omega)$, ground-state correlation function $\langle \hat{E}_{\textrm{vac},x}(\vr,\Omega) \hat{E}^\dagger_{\textrm{vac},x}(\rp,\Omega)\rangle$ and signal density $s^2(\vr,\rp,\Omega)  $ defined by $ \langle : \hat{S}^2 : \rangle =\int \dif \Omega \int \dif^3 r \int \dif^3 r^\prime s^2(\vr,\rp,\Omega) $ for two different cases: In the first row we plot them as functions of $x$ and $y$ by setting \mbox{$z=z^\prime=\rp_\parallel= 0$} and $\Omega =300\times 2\pi\,$TH whereas in the second row we set \mbox{$\vec{r}_\parallel = \rp_\parallel = z^\prime= 0$} such that the only free variables are $z$ and $\Omega$. %For the optical properties and approximations used to generate this Figure see Ref.~\cite{SUPP}. Note that all plots have been rescaled such that for each of them the maximal value is below one to ease the qualitative discussion.
 }
\label{Table}
\end{figure}

The structure of Eq.~\eqref{eq33:Scor2OfFilter} furnishes us with a clear physical picture for electro-optic sampling of vacuum fluctuations. The ground state correlation function of the electric field is sampled in a confined spatial region and a certain frequency interval defined by the spectral and spatial profile of the probe pulse. Which part of the correlation function is accessed can be adjusted by tuning the experimental parameters such as the pump pulse profile or properties of the crystal which in turn determine the filter function, as shown in Fig.~\ref{Table}. This flexibility means that electro-optic sampling represents a much more versatile experimental route to accessing the quantum vacuum compared to more well-established methods such as the Purcell effect (which only accesses the two-point correlation function in the coincidence limit) or the Casimir force (to which all frequencies contribute). Since we did not specify the laser pulse profile or the electromagnetic environment of the crystal and included absorption effects, Eq.~\eqref{eq33:Scor2OfFilter} can be used as a starting point for studying the structure of the medium-assisted quantum vacuum in general absorptive and dispersive environments targeted at chosen spectral and spatial regions. This allows one to study the polaritonic nature of the ground state inside the crystal with unprecedented versatility. In order to demonstrate the validity of the model, we first compare our result to previous theoretical and experimental works. \\

%We take a closer look at different approximated results for $\langle : \hat{S}^2 : \rangle$ derived from Eq.~\eqref{eq33:Scor2OfFilter}. Since we obtain in certain limits the result obtained previously in Ref.~\cite{Moskalenko2015} we are also able to compare our results to those of previous theoretical work. In order to simplify Eq.~\eqref{eq33:Scor2OfFilter} we can 
In order to make contact with the theory constructed in Ref.~\cite{Moskalenko2015}, one needs to assume the laser pulse to have a Gaussian profile %and apply a series of approximations to our general formula \eqref{eq33:Scor2OfFilter}. The first is to 
, neglect absorption and apply the paraxial approximation to the laser and vacuum fields by taking $k(\omega) \gg 1/\beamwaist$ then $ q(\Omega) \gg 1/\beamwaist$ where $k(\omega)$ and $q(\Omega)$ are the wave vectors of the laser and the vacuum, respectively. %Lastly, we can Taylor-expand the integrand found in Eq.~\eqref{eq33:Scor2OfFilter} in orders of $q_{z}/q$ up to second order which also allows one to perform the same amount of integrals analytically as in the case of the full paraxial approximation. For details, see Ref.~[PRA].Note that we obtain the result found in Ref.~\cite{Moskalenko2015} by neglecting absorption and applying the full paraxial approximation. \\
Such a procedure then reproduces precisely the result found in Ref.~\cite{Moskalenko2015}.
\begin{figure}
\centering
\includegraphics[width=1\columnwidth]{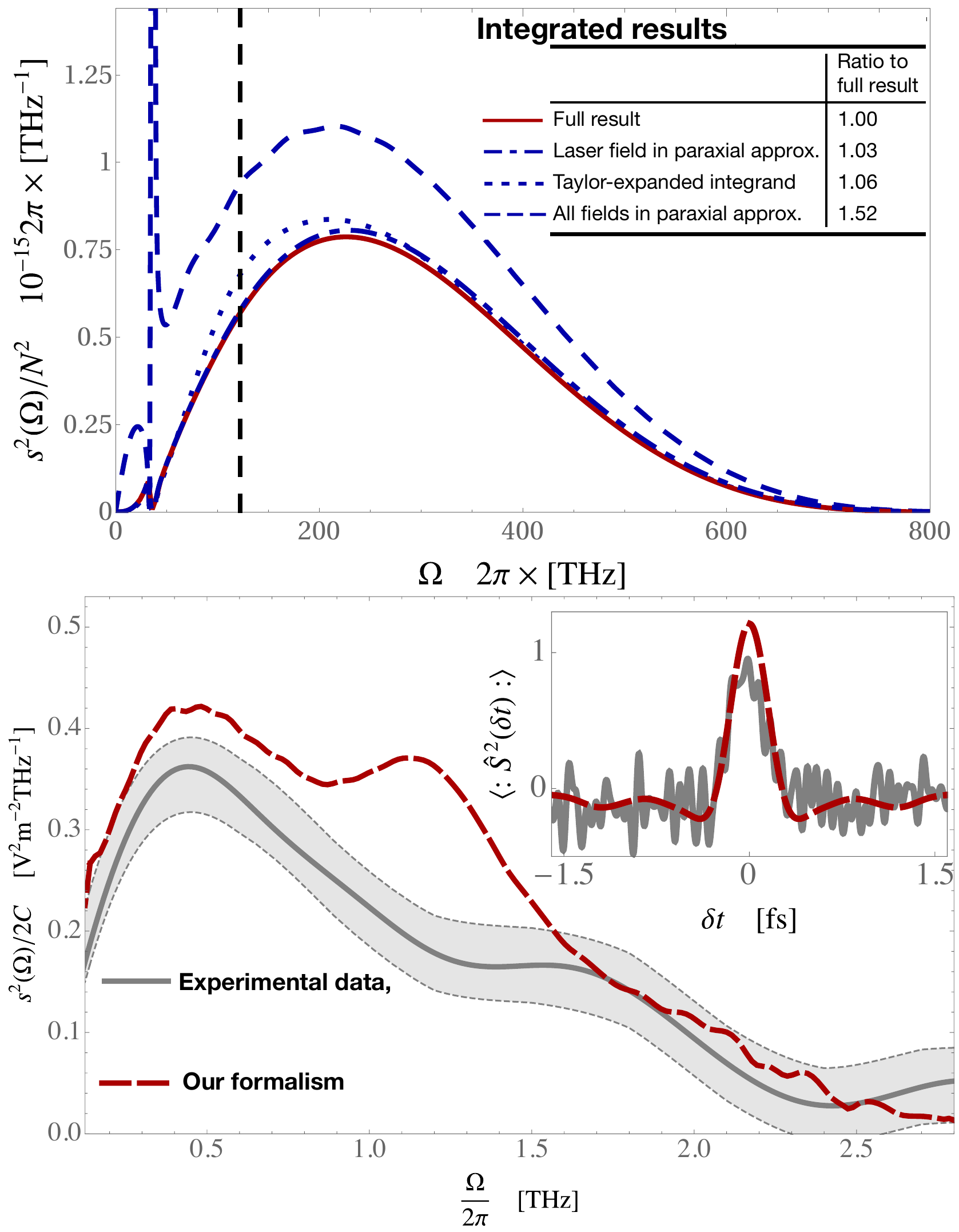}
\caption{\textit{Vacuum fluctuations in a bulk medium:} Upper plot: We plot $s^2(\Omega)/N^2$ without approximations (`Full result'), and in the different various approximations: laser paraxial, Taylor expanded integrand and the paraxial approximation. %The ratio of the total contribution to the variance $\langle : \hat{S}^2 : \rangle$ found using each approximation to our full result is shown in the legend.
 The cut-off discussed in the main text is shown by the dashed vertical line. Lower plot: We plot the signals spectrum $s^2(\Omega)$ normalised by $\sqrt{C} =2 \chi^{(2)} L \omega_p N/n \epsilon_0 c$ which is obtained from $\langle :\hat{S}^2(\delta t) : \rangle$ shown in the inset via a Fourier transformation. For the parameters used see supplementary materials \cite{SUPP}. The experimental data and its standard deviation taken from Ref.~\cite{Benea-Chelmus2019} are shown in gray. Note that we used a different convention for the Fourier transform in order to obtain the spectrum from the time domain data compared to Ref.~\cite{Benea-Chelmus2019}. }
\label{fig:Comparison}
\end{figure}
In order to assess the validity of the different approximations we use the same parameters as in Ref.~\cite{Moskalenko2015} (also listed in the Supplementary \cite{SUPP}), which were in turn realised experimentally in Ref.~\cite{Riek2015}. The result for the integrand $s^2(\Omega)$ defined by $\langle : \hat{S}^2 : \rangle = \int_0^\infty \dif \Omega s^2(\Omega)$ in case of the different approximations is shown in Fig.~\ref{fig:Comparison}. We find that in this parameter regime absorption can be neglected, since the frequency of the only relevant material resonance is well below the most relevant frequency range sampled in the experiment. However, we see that while the result with the paraxial approximation applied to the laser field agrees reasonably well with the full result obtained by direct evaluation of Eq.~\eqref{eq33:Scor2OfFilter}, applying the paraxial approximation to the vacuum field induces an error of $52\,\%$. Note that when follwoing the suggestion of the authors of Ref.~\cite{Moskalenko2015} of a cut-off of the signal's spectrum at $n(\Omega) \Omega < c\pi/\beamwaist$, the predicted integrated signal differs from our more complex theory by $12$\%. A good tradeoff between simplicity of expression and inclusion of all relevant physical effects is found by Taylor expanding the integrand to find a next-to-leading order paraxial approximation applied to the vacuum field which agrees within $6$\% of the full result, see supplementary materials for details \cite{SUPP}. \\

Next, we turn our attention to the parameter regime exploited in Ref.~\citep{Benea-Chelmus2019} where two spatially and temporally separated laser beams are used. Again, we can derive a filter function from first principles for this experimental setup which is done in Ref.~\cite{PRA} and the resulting expression together with the parameters under consideration can be found in the supplementary materials \cite{SUPP}. Strikingly, using two laser beams one can make a correlation measurement of the polaritonic ground state between different spatio-temporal regions. This allows one to obtain the spectrum $s^2(\Omega)$ %defined by $\langle : \hat{S}^2 :\rangle =\int_0^\infty \!\dif \Omega \, s^2(\Omega) $
 directly from the experimental data by Fourier transforming the measured signal $ \hat{S}^2(\delta t)$ obtained with different temporal shifts $\delta t$ between the laser pulses, i.e. \cite{PRA} $1/(2\pi)  \!\! \int_{-\infty}^\infty \!\dif \delta t \, \langle : \!\! \hat{S}^2(\delta t)\!\! : \rangle \me^{\mi \delta t \Omega }  = \frac{1}{2}s^2(|\Omega|) .$
 This key advantage enables one to study the quantum vacuum in greater detail, e.g. by accessing its spectrum. We apply the laser paraxial approximation and use the parameters experimentally realised in Ref.~\citep{Benea-Chelmus2019} which are summarised in the supplementary materials \cite{SUPP}. The result for the spectrum is compared to the experimental data in Fig.~\ref{fig:Comparison}. We find reasonable agreement between experiment and theory considering the errors on the input parameters. Note that our theoretical prediction does not contain any fitting parameter but is based on independently-measured optical properties such as the linear and nonlinear responds of the crystal. Furthermore, most of its contributions stem from a new term vanishing in the limit of vanishing absorption, as we show in detail in the supplementary materials \cite{SUPP}. This shows the importance of including absorption into the description of these experiments. Also note that in this parameter regime, one mainly accesses thermal fluctuations and not zero-point ones in contrast with \cite{Riek2015}. However, it is still a good test for our theory which treats zero-point and thermal fluctuations on the same footing (see Eq.~\eqref{eq:VacCorrelations}).\\\\
 \begin{figure}
\centering
\includegraphics[width=\columnwidth]{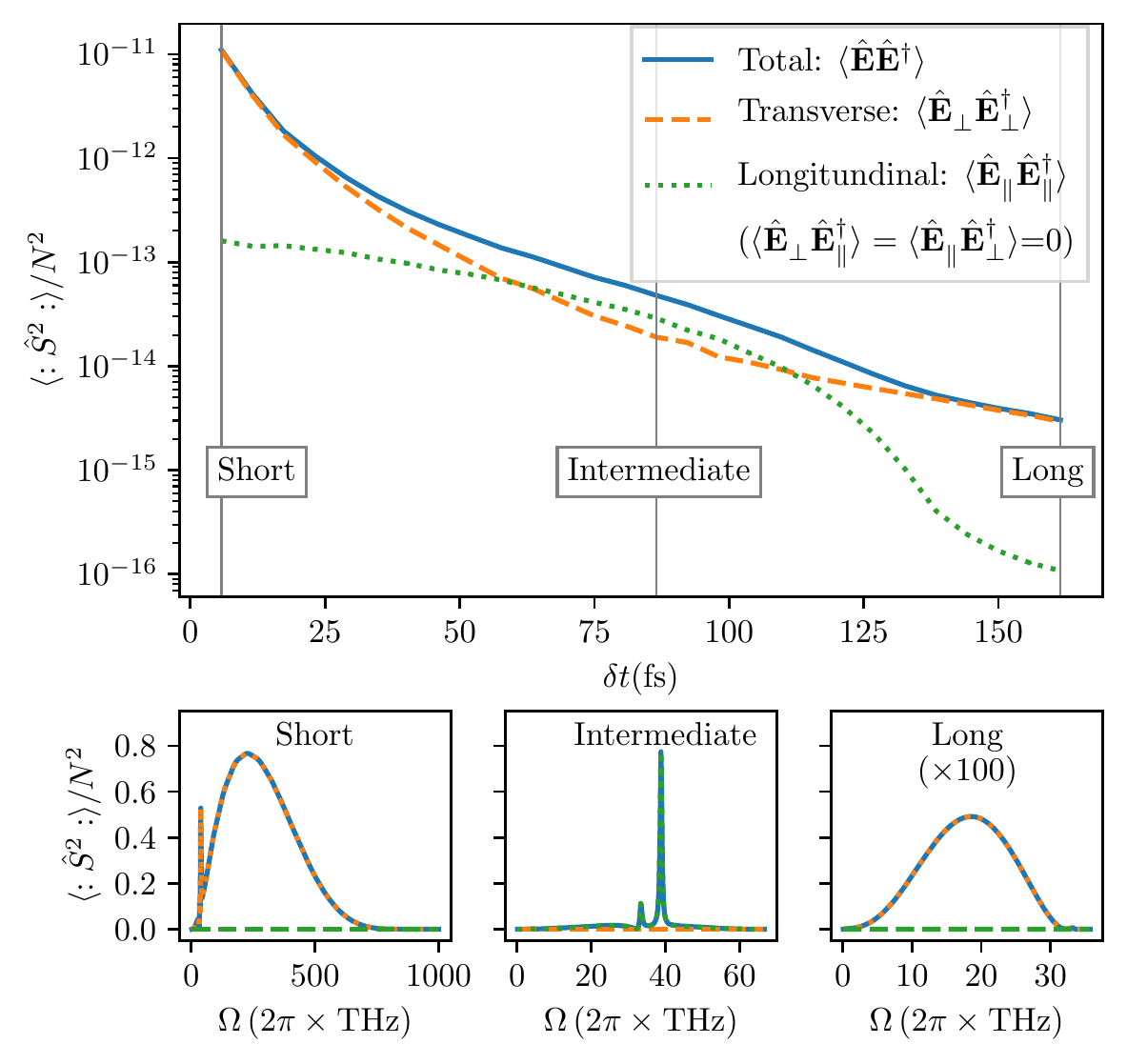}
\caption{\textit{Contribution to the variance for different pulse durations distinguishing matter and free field fluctuations:} We see in (a) that there is a intermediate interval of $\Delta t$ in which $\langle : \hat{S}^2 :\rangle $ is dominated by longitudinal contributions. This can be explained by considering the spectra plotted for different values of $\Delta t$ in (b), (c) and (d) as explained in the main text. }
\label{fig:Longitudinal}
\end{figure}
Having validated our new theoretical approach we are ready to exploit its full potential by revealing how electro-optic sampling experiments can be used to gain fundamental insight into the nature of the quantum vacuum inside the crystal, rather than simply describing its statistics. The ground state inside the crystal is that of the coupled system of the electromagnetic field and the charges. Hence, the ground state fluctuations consist of those of both the `free' field and the near-field generated by the fluctuating charges inside the crystal --- in the following referred to as matter fluctuations. It is important to notice that what we call `free' field is not the same as the fluctuating field in empty space, but rather the photon-like part of the interacting system of photon and charges. In Coulomb gauge one can distinguish the two different types of contributions to the quantum vacuum of the electromagnetic field by decomposing the electric-field operator into its longitudinal ($\parallel$) and transverse ($\perp$) components \cite{Philbin2010}.
 Using this in Eq.~\eqref{eq33:Scor2OfFilter}, we find contributions to the signal's variance stemming from free field and from matter fluctuations allowing one to analyze which of the two is accessed in the experiments. We use the same parameters as in Refs.~\cite{Moskalenko2015,Riek2015} except that we vary the pulse duration $\Delta t$ as shown in Fig.~\ref{fig:Longitudinal}. We find that in the parameter regime of Ref.~\cite{Moskalenko2015,Riek2015} where $\Delta t =5.9 $\,fs only transverse and hence free-field fluctuations contribute to the signal and the detected fluctuating field is dominated by photon-like fluctuations. This can be explained by the fact that the main frequency range which is resolved is far from any material resonances, c.f. \mbox{Fig.~\ref{fig:Longitudinal}(b)}. One can show analytically that the longitudinal part is proportional to Im$[\epsilon]$ such that far from material resonances only transverse fluctuating fields contribute \cite{SUPP}. \\
The situation changes for an intermediate pulse duration, where the resolved frequency range coincides with a material resonance, compare \mbox{Fig.~\ref{fig:Longitudinal}(c)}. This leads to the detection of polaritonic modes which are dominated by their matter content resulting in mainly longitudinal fluctuations. For even longer pulse duration, only field fluctuations well below the material resonance are detected, leading to a signal which is dominated by transverse fluctuating fields as indicated in \mbox{Fig.~\ref{fig:Longitudinal} (d)}. This analysis reveals the possibility to unambiguously interpret and identify different properties of the richly structured polaritonic quantum vacuum inside the crystal using the formalism developed here.  \\
  
In conclusion, we have outlined a theoretical framework for analysing and interpreting the quantum-vacuum detector as provided by electro-optic sampling experiments sensitive to the QED vacuum. Our model includes absorption effects, goes beyond the paraxial approximation and takes the full medium-assisted or polaritonic ground state into account. It agrees well with experimental data and offers significant improvements on previous theoretical works in an experimentally-realised parameter regime. In addition, it provides a starting point for a detailed analysis of the quantum vacuum in media and its rich structure in new, so far theoretically inaccessible, parameter regimes. As an example, it was shown that transverse and longitudinal fluctuating field can be analysed individually, revealing the polaritonic nature of the QED ground state in media. Other characteristics of the quantum vacuum might be accessible using electro-optic sampling such as the influence of additional surfaces onto the electromagnetic ground state which is of relevance to e.g. the Purcell or Casimir effect, or adhesion forces. Apart from electro-optic sampling, the general formalism resulting from our combining of macroscopic QED with nonlinear optics has applications in a wide range of fields such as recent studies of analogues of the dynamical Casimir effect \cite{Vezzoli2018}, pair generation in $\epsilon$-near zero material or metamaterials \cite{prain2017spontaneous} and photonic Bose-Einstein condensates \cite{nyman2014interactions}.\\

\acknowledgments{ The authors thank Stephen Barnett, Thomas Wellens, Vyacheslav Shatokhin, Giacomo Sorelli, Niclas Westerberg, Christoph Dittel, Jerome Faist, Ileana-Cristina Benea-Chelmus, Francesca Fabiana Settembrini, Denis Seletskiy, Guido Burkard and Alfred Leitenstorfer, for fruitful discussions. R.B. acknowledges financial support by the Alexander von Humboldt Foundation, S.Y.B. thanks the Deutsche Forschungsgemeinschaft (grant BU 1803/3-1476). R.B. and S.Y.B. both acknowledge support from the Freiburg Institute for Advanced Studies (FRIAS).}

\newpage

\begin{widetext}

\begin{center}
\textbf{\large Supplementary material}
\end{center}

\section{Green's tensor}

The Green's tensor of the vector Helmholtz equation is defined by 
\begin{align} \label{eq:GreensTensorDef}
 \nabla \times \nabla \times  \tens{G}(\vec{r},\vec{r}^\prime,\omega) - \frac{\omega^2}{c^2} \epsilon(\vec{r},\omega)   \tens{G}(\vec{r},\vec{r}^\prime,\omega) = \Greektens{$\delta$}(\vec{r} - \vec{r}^\prime )
\end{align}
and the boundary condition $\tens{G}(\vec{r},\vec{r}^\prime,\omega) \to 0$ for $|\vec{r} -\vec{r}^\prime| \to \infty$. Here, $\epsilon$ is the permittivity. In a $(2+1)$ dimensional Weyl decomposition it is given by \cite{Buhmann2012a}
\begin{multline} \label{eq:BulkGreensTensor}
\tens{G}^{(0)}(\vec{r},\vec{r}^\prime, \omega) = - \frac{1}{4\pi^2 k^2(\omega)} \int \dif^2 k_\parallel \, \frac{\me^{i\vec{k}_\parallel \cdot (\vec{r}-\vec{r}^\prime)}}{\kPerp}\delta(z - z^\prime) \vec{e}_z \vec{e}_z
+ \frac{i}{8\pi^2} \int \dif^2 k_\parallel \frac{\me^{i\vec{k}_\parallel \cdot (\vec{r}-\vec{r}^\prime)}}{k_z} \\
\times \sum_{\sigma=s,p} \left[ \vec{e}_{\sigma+}\vec{e}_{\sigma+} \me^{i k_z(z-z^\prime)} \theta(z - z^\prime) +\vec{e}_{\sigma-}\vec{e}_{\sigma-} \me^{-i k_z(z-z^\prime)} \theta(z^\prime - z)\right].
\end{multline}
Here $k =\sqrt{\epsilon(\omega)} \omega/c$, and we have $k_z = k_z(k_\parallel, \omega)=\sqrt{k^2-k_\parallel^2}$ with $\textrm{Im}[k_z]>0$. Furthermore, the polarisation vectors $\vec{e}_{\sigma\pm}$ with $\sigma = s,p$ read
\begin{align} \label{eq:GreensTensorPolarizationVectors}
\vec{e}_{s\pm}  =  \frac{1}{k_\parallel} \left( \begin{array}{c}
 k_y \\ - k_x \\ 0 
\end{array} \right);   \quad \quad  \vec{e}_{p\pm}  = \frac{1}{k} \left(
\begin{array}{c}
 \mp \frac{k_x k_z}{k_\parallel} \\ \mp \frac{k_y k_z}{k_\parallel} \\ k_\parallel
\end{array} \right) .
\end{align}
In order to distinguish longitudinal and transverse quantum fluctuations, one needs the transverse ($^{\perp}\tens{G}^{(0)}$) and longitudinal ($^{\parallel}\tens{G}^{(0)}$) part of the Green's tensor defined by 
\begin{align} \label{eqapp:DefinTrans}
^{\perp}\tens{G}(\vec{r},\vec{r}^\prime) \equiv \int \dif^3 s \,\,\, \Greektens{$\delta$}^{\perp }(\vec{r}-\vec{s}) \cdot \tens{G}(\vec{s},\vec{r}^\prime) ,\\ \label{eqapp:DefinLong}
^{\parallel}\tens{G}(\vec{r},\vec{r}^\prime) \equiv \int \dif^3 s \,\,\, \Greektens{$\delta$}^{\parallel}(\vec{r}-\vec{s}) \cdot \tens{G}(\vec{s},\vec{r}^\prime).
\end{align}
In Fourier space we have
\begin{align}
\Greektens{$\delta$}^\perp(\vec{r}) & = \frac{1}{(2\pi)^3} \int \dif^3 k \,\, \me^{\mi \vec{k}\cdot \vec{r}} \left( \tens{I} - \vec{e}_\vec{k}\vec{e}_\vec{k}     \right), \\ \label{eqapp:DeltaParallel}
\Greektens{$\delta$}^\parallel(\vec{r})& = \frac{1}{(2\pi)^3} \int \dif^3 k \,\, \me^{\mi \vec{k}\cdot \vec{r}} \vec{e}_\vec{k}\vec{e}_\vec{k},
\end{align}
where $\tens{I}$ is the unit tensor. Inserting Eq.~\eqref{eqapp:DeltaParallel} into Eq.~\eqref{eqapp:DefinLong}, some algebra shows
\begin{multline} \label{eq:Gparallel}
^{\parallel}\tens{G}^{(0)}(\vec{r},\vec{r}^\prime) = - \frac{1}{4\pi^2 k^2} \int \dif^2 k_\parallel \me^{\mi \vec{k}_\parallel\cdot (\vec{r}-\vec{r}^\prime)}\delta(z-z^\prime) \vec{e}_z \vec{e}_z 
- \frac{1}{8\pi^2 k^2} \int \dif^2 k_\parallel \me^{\mi \vec{k}_\parallel \cdot (\vec{r}-\vec{r}^\prime)} \me^{-k_\parallel | z-z^\prime|} \\ \times \left[ k_\parallel (\vec{e}_{k_\parallel} \vec{e}_{k_\parallel} -\vec{e}_z \vec{e}_z) + \mi \,\, \textrm{sign}(z-z^\prime)  \left( \hspace{-5pt} \begin{array}{ccc} 0 & 0 & k_x \\
0 & 0 & k_y \\
k_x & k_y & 0
\end{array}  \hspace{-5pt}  \right)       \right].
\end{multline}
Having found the longitudinal part of the Green's tensor its transverse part is simply given by $ ^{\perp}\tens{G}(\vec{r},\vec{r}^\prime) =\tens{G}(\vec{r},\vec{r}^\prime) -  ^{\parallel}\tens{G}(\vec{r},\vec{r}^\prime)$.\\
Note that Onsager reciprocity $\tens{G}^{(0)}(\vec{r},\rp)=\tens{G}^{^{(0)}\text{T}}(\rp,\vec{r})$ holds independently for the longitudinal and transverse parts of the Green's tensor, \mbox{$^{\parallel (\perp)}\tens{G}^{(0)}(\vec{r},\vec{r}^\prime) = \tens{G}^{^{(0)}\text{T}\parallel (\perp)}(\rp,\vec{r}$)}  \cite{Buhmann2012a}. Further, Eq.~\eqref{eq:Gparallel} shows that the longitudinal component of the Green's tensor is a symmetric tensor and is also symmetric under the exchange of it position arguments. This implies $^{\parallel}\tens{G}^{(0)}(\vec{r},\vec{r}^\prime) =\tens{G}^{(0)}(\vec{r},\vec{r}^\prime)^\parallel $. From hence in turn, we see that \mbox{$ ^{\parallel(\perp)}\tens{G}^{(0)}(\vec{r},\vec{r}^\prime)^{\parallel(\perp)}=  ^{\parallel(\perp)}\!\!\tens{G}^{(0)}(\vec{r},\vec{r}^\prime)$} and
\begin{align} \label{eq:GperpGpar}
^{\parallel(\perp)}\tens{G}^{(0)}(\vec{r},\vec{r}^\prime)^{\perp(\parallel)} = ^{\perp(\parallel)}\!\!\tens{G}^{(0)}(\vec{r},\vec{r}^\prime)^{\parallel(\perp)}  = \mathsf{\mathbf{0}}.
\end{align}

\section{Filter function and signal}

We state the formulae for the electro-optical signal and the filter function using different approximations as described in the main text. For a more detailed derivation see Ref.~\cite{PRA}. \\
The full filter function which can be derived from first principles based on the formalism developed here reads \cite{PRA}
\begin{multline}\label{eq33:FilterOfH}
F(\vec{r},\vec{r}^\prime,\Omega,\Omega^\prime) = \frac{1}{2}\left\{\left[  H_1(\vec{r},\Omega) + H_1^\ast(\vec{r},-\Omega) \right] \left[ H_2(\vec{r}^\prime,-\Omega^\prime) + H_2^\ast(\vec{r}^\prime,\Omega^\prime) \right] \right. \\ \left. + \left[  H_2(\vec{r},\Omega) + H_2^\ast(\vec{r},-\Omega) \right] \left[ H_1(\vec{r}^\prime,-\Omega^\prime) 
 + H_1^\ast(\vec{r}^\prime,\Omega^\prime) \right] \right\}.
\end{multline}
where 
\begin{align}\label{eq:hfunction}
H_i(\vec{r}^\prime,\Omega)  = -8 \pi \mi  c \epsilon_0  \chi^{(2)}(\Omega) \mu_0 \int\limits_0^\infty \dif\omega \frac{\eta(\omega) \sqrt{\epsilon(\omega)} \omega}{\hbar}
 \int \dif^2 r_\parallel E_{i,y}^\ast(\vec{r}_\parallel,\omega) \textsf{G}_{xx} (\vec{r}_\parallel,\vec{r}^\prime, \omega)E_{i,y}(\vec{r}^\prime, \omega-\Omega).
\end{align}
Here the two laser pulses are defined by
\begin{subequations}\label{eq33:TwoBeamE1E2Def}
\begin{align} 
E_{1,y}(\vec{r},t) & = \int_{-\infty}^\infty \dif t \,\, E_{1,y}(\vec{r},\omega) \me^{\mi \omega t}, \\
E_{2,y}(\vec{r},t) & = \int_{-\infty}^\infty \dif t \,\, E_{1,y}(\vec{r}+\delta \vec{r},\omega) \me^{\mi \omega \delta t}  E_\delta(\vec{r},\omega) \me^{\mi \omega t}\vec{e}_y \nonumber \\
 & \equiv \int_{-\infty}^\infty \dif t \,\,  E_{2,y}(\vec{r},\omega) \me^{\mi \omega t}\vec{e}_y.
\end{align}
\end{subequations}
This implies \mbox{$E_{2,y}(\vec{r},\omega) \equiv E_{1,y}(\vec{r}+\delta \vec{r},\omega) \me^{\mi \omega \delta t}$}. In the following we only consider the single-beam setup obtained in the limit $\delta t=0$ and $ \delta \vec{r}_\parallel = 0$. \\
 In a first step we neglect absorption by assuming that the refractive index $n$ is real for all frequencies under consideration. This is a reasonable approximation in the parameter range considered in the upper plot in Fig.~3. Furthermore, we assume that the laser pulse is a Laguerre-Gaussian mode of lowest order and that the crystal length $L$ is much shorter than the Rayleigh length of the beam, i.e. $\beamwaist^2 k /2 \gg L $, such that the beam is given by
 \begin{align}\label{eq24:PulseFundamentalGaussian}
\Eprw = E_\textrm{p}(\omega) \sqrt{\frac{2}{\pi \beamwaist^2}} \me^{- r_\parallel^2/\beamwaist^2} \me^{\mi k z} \vec{e}_y.
\end{align}
Inserting Eqs.~\eqref{eq33:FilterOfH} and Eq.~\eqref{eq24:PulseFundamentalGaussian} into Eq.~(4) one obtains the full result 
 \begin{multline} \label{eqapp:FullResult}
\langle :\hat{S}^2 : \rangle = \frac{\hbar \mu_0 \beamwaist^4 L^2}{4\pi^3 c^2} \int_0^\infty \dif \Omega \, |\chi^{(2)}(\Omega)|^2 \Omega^2 \,\, ¸\mathclap{\int\limits_{q_\parallel < q}} \,\,\,\, \dif^2 q_\parallel \, \left(1- \frac{q_x^2}{q^2}\right)\frac{ \me^{- q_\parallel^2 \beamwaist^2/2}}{q_z}    \bigg|   \int_0^\infty \dif \omega \int \dif^2 k_\parallel \,\frac{n(\omega)\eta(\omega)\omega}{\hbar^2} \left(1- \frac{k_x^2}{k^2} \right) \\
 \times\me^{-(k_\parallel^2 + \vec{q}_\parallel \cdot \vec{k}_\parallel ) \beamwaist^2/2} \left[  E^{\ast}_\textrm{p}(\omega) E_\textrm{p}(\omega - \Omega) \frac{\textrm{sinc}[\frac{L}{2} \Delta K_-] }{k_z} +  c.c.(\Omega \to -\Omega) \right] \bigg|^2 +(\Delta K_- \to \Delta K_+).
\end{multline}
Here we have defined $ \Delta K_\mp = k_{\omega-\Omega}-k_z \pm q_z$, and $+(\Delta K_- \to \Delta K_+)$ refers to the same term subject to the replacement $\Delta K_- \to \Delta K_+$. \\
In laser paraxial approximation, i.e. assuming $n^2(\omega) \omega^2/c^2 \gg \frac{1}{\beamWaist^2} $ within the frequency range of the laser one obtains
\begin{align} \label{eq33:S2PPFinal}
s^2(\Omega) = \frac{\left(N  L \omega_p \right)^2 \hbar |\chi^{(2)}(\Omega)|^2 }{4 \pi^3 c^4 \epsilon_0^3 n^2(\omega_c)}   \Omega^2 f(\Omega)^2 \!\!\! \int\limits_{\mathclap{q_{\parallel} \leq q}} \dif^2 q_{\parallel} \frac{1-\frac{q_x^2}{q^2}}{q_z}  \me^{-q_{\parallel}^2 \beamWaist^2/4}
 \left( \text{sinc} \left[\frac{L}{2} \Delta k_- \right]^2 +\text{sinc} \left[\frac{L}{2}\Delta k_+ \right]^2 \right).
\end{align}
Here, we have introduced the shorthand $\Delta k_\pm = n_g \frac{\Omega}{c} \pm q_z$ where $q_z = \sqrt{q^2 - q_{\parallel}^2 }$ and $q=n(\Omega)\Omega/c$ is the wave vector at the frequency of the vacuum $\Omega$. Furthermore $n_g = c \partial k/\partial \omega |_{\omega=\omega_c}$ is the group refractive index at the central frequency of the laser pulse $\omega_c$ and we have defined 
\begin{align}  \label{eq33:Omegap}
\omega_p & =  \int_0^\infty \!\!\!\! \dif \omega \eta(\omega) E_\textrm{p}(\omega)^2 \bigg/\int_0^\infty \!\!\!\! \dif \omega \frac{\eta(\omega)}{\omega} E_\textrm{p}(\omega)^2,  \\ \label{eq33:SpectralAutocorrlation}
f(\Omega)& = \frac{ \int_0^\infty  \dif \omega \left[ E_\textrm{p}(\omega) E_\textrm{p}(\omega + \Omega) + E_\textrm{p}(\omega) E_\textrm{p}(\omega - \Omega)  \right]}{ 2\int_0^\infty \dif \omega \eta(\omega) E_\textrm{p}(\omega)^2}, \\  \label{eq33:N}
N & = 4 \pi \epsilon_0 c n(\omega_c) \int_0^\infty \dif \omega \frac{\eta(\omega)}{\hbar \omega} E_\textrm{p}^2(\omega). 
\end{align}
Taylor-expanding Eq.~\eqref{eq33:S2PPFinal} up to fourth order in $q_\parallel/q$ one finds
\begin{multline} \label{eq33:S2TaylorExpanded}
s^2(\Omega)= \frac{(N  L \omega_p)^2 \hbar}{\pi^2 \epsilon_0^3 c^3 n^3(\omega_c) \textrm{w}^2}  \, \frac{n(\omega_c)}{n(\Omega)} |\chi^{(2)}(\Omega)|^2 \Omega  f(\Omega)^2 \left\{   \left( 1- \me^{-q^2 \textrm{w}^2/4} \right) %\left(
 \text{sinc} \left[ \frac{L}{2} \Delta k_- \right]^2 
% +\text{sinc} \left[\frac{L}{2} \Delta k_- \right]^2  \right) 
+ (\Delta k_- \to \Delta k_-)  \right. \\ \left.
- \left[  \frac{4 - \me^{-q^2 \omega_0^2/4} (4 + q^2 \omega_0^2)}{q  \omega_0^2} \left( \Delta k_- \text{sinc} \left[ L \Delta k_- \right]  -  \Delta k_+ \text{sinc}\left[ \frac{L}{2} \Delta k_- \right]^2  + (\Delta k_- \to \Delta k_+) 
%+\Delta k_- \text{sinc} \left[ L \Delta k_- \right]  -   \Delta k_- \text{sinc}\left[ \frac{L}{2} \Delta k_- \right]^2
\right)   \right]  \right\}.
\end{multline}
Only considering the lowest order in the Taylor expansion in Eq.~\eqref{eq33:S2TaylorExpanded} one finds the full paraxial approximation 
\begin{align} \label{eq33:S2ParaxialFinal}
s^2(\Omega) = \frac{(N  L \omega_p)^2 \hbar}{\pi^2 \epsilon_0^3 c^3 n^3(\omega_c) \textrm{w}^2} \frac{n(\omega_c)}{n(\Omega)} |\chi^{(2)}(\Omega)|^2 \Omega f(\Omega)^2 
 \left( \text{sinc} \left[\frac{L}{2}\Delta k_- \right]^2 +\text{sinc} \left[\frac{L}{2}\Delta k_+ \right]^2 \right).
\end{align} 
Eqs.~\eqref{eqapp:FullResult}, \eqref{eq33:S2PPFinal}, \eqref{eq33:S2TaylorExpanded} and \eqref{eq33:S2ParaxialFinal} where used to generate the upper plot in Fig.~3 in the main text. Eq.~\eqref{eq33:S2PPFinal} has been used to generate the plots in Fig.~2.\\
Not neglecting absorption by allowing the refractive index $n$ to be a complex quantity but by applying the laser-paraxial approximation one finds
             \begin{multline}\label{eq33:sOmegaAbs}
s^2(\Omega)= \hbar \frac{\left(N \omega_p   \Omega f(\Omega) |\chi^{(2)}(\Omega)| \right)^2  [2 n_T(\Omega) + 1]}{2 \pi^2 c^4 \epsilon_0^3 n^2}  \int\limits_0^\infty \dif q_\parallel\,\, q_\parallel  \me^{-q_\parallel^2 \textrm{w}^2/4} \\
\times \text{Re}\left[\left(2 - \frac{q_\parallel^2 }{q^2} \right)\left( \frac{-\mi L}{  q_z (n_g \Omega/c - q_z)} + \frac{1 - \me^{\mi L (q_z - n_g\Omega/c)}}{  q_z (q_z - n_g\Omega/c)^2 } \right) + n_g \to -n_g \right].
\end{multline}
This equation has been used to generate the lower plot Fig.~3. \\

In order to distinguish longitudinal and transverse fluctuations we calculate the signal stemming from longitudinal fluctuations only defined by 
\begin{align} \label{eq:Spar} 
\langle : \hat{S}^2_\parallel : \rangle = \int_0^\infty \!\! \dif \Omega \,\, s_\parallel^2(\Omega) 
= \intzi \dif \Omega \intzi \dif \Omega^\prime \int_{V_\textrm{C}} \dif^3 r^\prime \int_{V_\textrm{C}} \dif^3 r^{\prime\prime} \langle ^\parallel E_{\textrm{vac},x}(\vec{r}^\prime, \Omega)  E^{\parallel \dagger}_{\textrm{vac},x}(\vec{r}^{\prime\prime}, \Omega^\prime) \rangle  F(\vec{r}^\prime, \vec{r}^{\prime\prime},\Omega,\Omega^\prime).
\end{align}
Inserting Eqs.~(5) and \eqref{eq:Gparallel} into Eq.~\eqref{eq:Spar} we obtain
\begin{multline} \label{eq44:sparallel}
 s^2_\parallel(\Omega)  = \left( \frac{N \omega_p }{n c}    \right)^2 \frac{\hbar|\chi^{(2)}(\Omega)|^2 f(\Omega)^2 \textrm{Im}\left[\epsilon(\Omega)\right] }{\epsilon_0^3 \pi^3 |n(\Omega)|^4}  \left\{  \frac{2L}{\omega_0^2} \right. \\ \left.  - \frac{L n_g^2 \Omega^2 \me^{[n_g \Omega\omega_0/(2c)]^2}}{2 c^2}\Gamma\left(\frac{n_g^2 \Omega^2 \omega_0^2}{4c^2}\right) \! + \! \int_0^\infty \!\!\!\dif k_\parallel \, k_\parallel^2 \me^{-k_\parallel^2 w_0^2/4}\textrm{Re}\left[ \frac{\me^{-(k_\parallel +\mi n_g \Omega/c)L}-1}{(k_\parallel + \mi n_g \Omega/c)^2}      \right]  \right\}.
\end{multline}
Here, \mbox{$\Gamma(z) = \int_z^\infty\dif z\, \me^{-z} z^{-1}$} is the incomplete Gamma function. Since $\langle \hat{\vec{E}} \hat{\vec{E}}^\dagger \rangle \propto \textrm{Im} \tens{G}$, it follows directly from Eq.~\eqref{eq:GperpGpar} that longitudinal and transverse fluctuations are uncorrelated, i.e. $\langle^\perp  \hat{\vec{E}} \hat{\vec{E}}^{\parallel \dagger} \rangle =\langle^\parallel  \hat{\vec{E}} \hat{\vec{E}}^{\perp\dagger} \rangle =0$. So we obtain the signal originating from transverse fluctuations via $s^2_\perp(\Omega)= s^2(\Omega) - s^2_\parallel(\Omega)$.

\section{Parameters used}

In order to generate the upper plot in Fig.~3 and Fig.~2 we used the following parameters also considered in Ref.~\cite{Moskalenko2015}. The crystal length is given by $L= 7\mu$m. The spectral decomposition of the laser pulse $E_\textrm{p}(\omega)$ has a rectangular shape being equal to one for $\omega \in [\omega_c -\Delta \omega/2, \omega_c + \Delta \omega/2]$ and zero elsewhere.  Here $\omega_c = 255 (2\pi)$THz and $\Delta \omega =75(2\pi) $THz. Furthermore, the beam waist is given by $\beamwaist = 3\mu$m. Concerning the optical properties of the ZnTe crystal we use that the nonlinear susceptibility is well approximated by a constant value over the full frequency range under consideration which is given by $\chi^{(2)} = 1.17 \times 10^{-21}\,\textrm{CV}^{-2}$ \cite{Leitenstorfer1999}. 
For the refractive index in the frequency range of the laser we use \cite{marple_refractive_1964}
\begin{align} \label{eq33:Refractive1}
n(\omega)^2 = A + \left(\frac{B \lambda^2}{\lambda^2 - C} \right),
\end{align}
Here we have $\lambda = 2\pi c/\omega$, $A=4.27$, $B = 3.01$ and $C=0.142$ leading to $n_g = 2.24$. In the THz frequency range the refractive index is modeled by \cite{Leitenstorfer1999}:
\begin{align}\label{eq33:Refractive2}
n(\Omega) = \textrm{Re}\left[ \sqrt{\epsilon_\infty \left( 1 + \frac{(\hbar \omega_\textrm{LO})^2 -(\hbar \omega_\textrm{TO})^2}{(\hbar \omega_\textrm{TO})^2 - (\hbar \Omega)^2 - \mi \hbar \gamma \Omega} \right)}      \right],
\end{align}
with $\omega_\textrm{TO} = 5.31 \times 2\pi \textrm{THz}$, $\omega_\textrm{LO} = 6.18  \times 2\pi \textrm{THz}$, $\gamma = 0.09 \times 2\pi\textrm{THz}$ and $\epsilon_\infty = 6.7$. Note that taking the real part in Eq.~\eqref{eq33:Refractive2} indicates that we neglect absorption effects.  \\
To generate the lower plot in Fig.~3 we chose the parameters according to Ref.~\cite{Benea-Chelmus2019}. Here we have $L = 3\,$mm and $\beamwaist = 125\mu $m. The spectral decomposition of the laser is given by $\Epw   = \sqrt{\frac{1}{\sigma }} \frac{\me^{-(\omega-\omega_c)^2/\sigma^2}}{(2\pi)^{1/4}} $, with  $\omega_c = 375\times (2\pi)$\,THz, $\sigma = \sqrt{2/\pi}(\Delta t)^{-1}$ and $\Delta t =80\, $fs. In the frequency range of the laser, we use the same refractive index also used to generate the lower plot in Fig.~3. Concerning the nonlinear susceptibility we use
\cite{Leitenstorfer1999}
\begin{align}\label{42:NonlinearSusDisp}
\chi^{(2)}(\Omega) = \frac{n^4(\omega_\textrm{C})\epsilon_0}{2} r_{41} \left[1 +C_0 \frac{(\hbar \omega_\textrm{TO})^2}{\hbar \Omega - \mi \hbar \Omega \gamma} \right],
\end{align}
with $\omega_\textrm{TO} = 5.31 \times 2\pi \textrm{THz}$, $\gamma = 0.09 \times 2\pi\textrm{THz}$, $C_0 = -0.07$, $r_{41}=1.17\times 10^{-21}\,\textrm{C}\textrm{V}^{-2}$. For the refractive index in the frequency range of the detected quantum vacuum we use the measurement data of the reals part of the refractive index  and the absorption coefficient $\alpha$. From the latter one obtains the imaginary part of the refractive index via $\alpha(\Omega)= \textrm{Im}[n(\Omega)] \Omega/c$. The real and imaginary parts of the refractive index retrieved from the measurement data is shown in Fig.~1.

\begin{figure}
\centering
\includegraphics[width=0.4\textwidth]{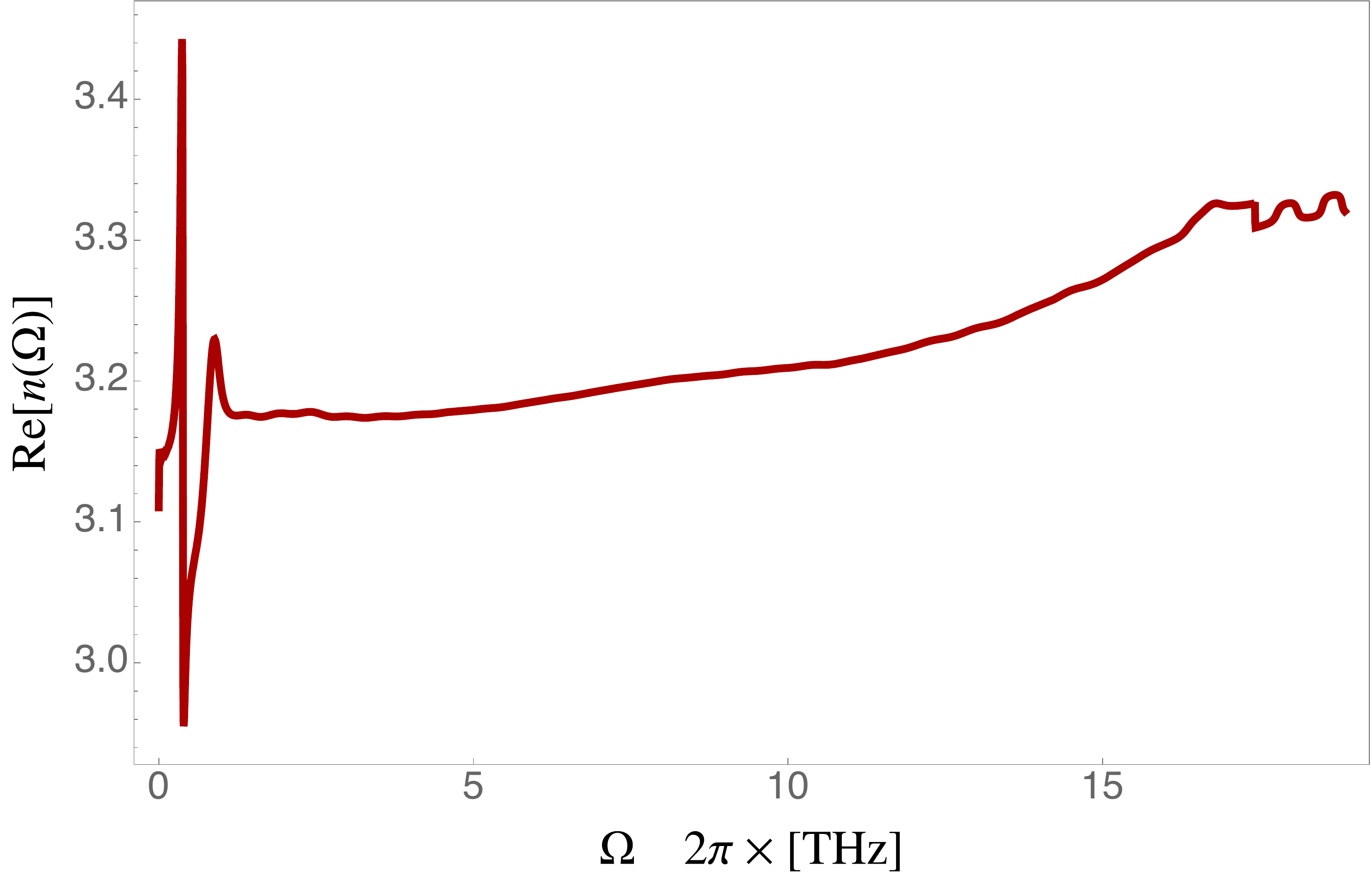} \hspace{1.5cm}
\includegraphics[width=0.4\textwidth]{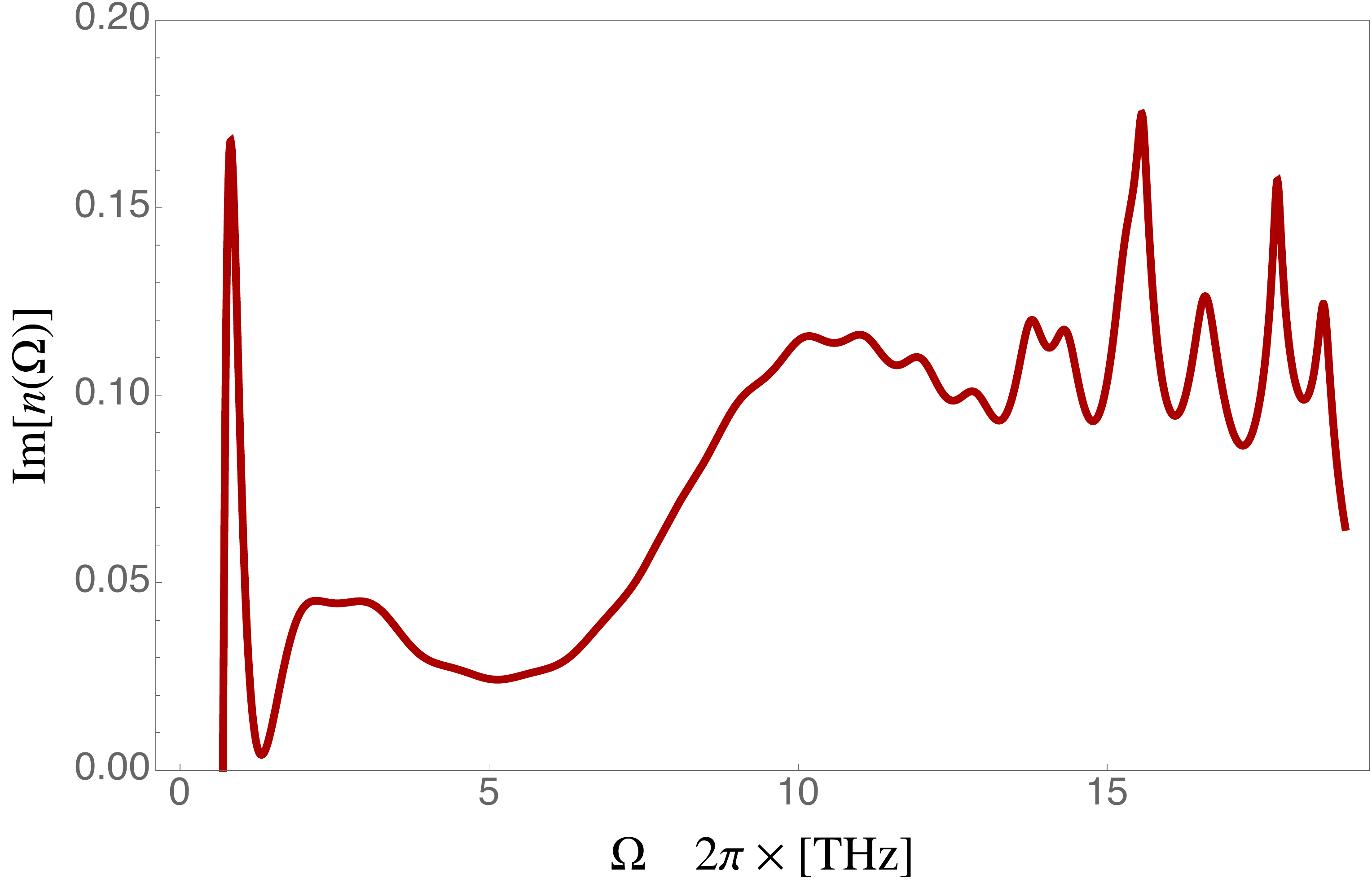}
\label{fig:Refractive}
\caption{ \textit{Real and imaginary parts of the refractive index:} Here the real and imaginary part of the refractive index at temperature $T=300$\,K as measured in Ref.~\citep{Benea-Chelmus2019} is plotted as a function of frequency. Note that while the real part of $n$ has been measured directly, the imaginary part of $n$ is calculated from the experimental data for the absorption coefficient $\alpha(\Omega)$ via $\alpha(\Omega) = \text{Im}[n(\Omega)]\Omega/c$. }
\end{figure}

\section{Origin of fluctuations accessed in Ref.~\cite{Benea-Chelmus2019}}
%\bibliographystyle{unsrt}
%\bibliography{library}
%\gdef\thesection{M\arabic{section}}
%\gdef\thesubsubsection{M\arabic{section}.\arabic{subsubsection}}
In order to compare our theory to the experimental data found in Ref.~\cite{Benea-Chelmus2019} as was shown in the bottom of Fig.~3 in the main text, we used Eq.~\eqref{eq33:sOmegaAbs}. Neglecting the off-resonant terms with $n_g \to -n_g$, this equation contains two different terms
\begin{align}  \label{eq:First}
s_1^2(\Omega) &= \hbar \frac{\left(N \omega_p   \Omega f(\Omega) |\chi^{(2)}(\Omega)| \right)^2  [2 n_T(\Omega) + 1]}{2 \pi^2 c^4 \varepsilon_0^3 n^2} 
\int\limits_0^\infty \dif q_\parallel\,\, q_\parallel  \me^{-q_\parallel^2 \textrm{w}^2/4} \text{Re}\Bigg[\left(2 - \frac{q_\parallel^2 }{q^2} \right) \frac{-\mi L}{  q_z (n_g \Omega/c - q_z)}  \Bigg],\\  \label{eq:Second}
s_2^2(\Omega) & = \hbar \frac{\left(N \omega_p   \Omega f(\Omega) |\chi^{(2)}(\Omega)| \right)^2  [2 n_T(\Omega) + 1]}{2 \pi^2 c^4 \varepsilon_0^3 n^2} 
\int\limits_0^\infty \dif q_\parallel\,\, q_\parallel  \me^{-q_\parallel^2 \textrm{w}^2/4} \text{Re}\Bigg[\left(2 - \frac{q_\parallel^2 }{q^2} \right)  \frac{1 - \me^{\mi L (q_z - n_g\Omega/c)}}{  q_z (q_z - n_g\Omega/c)^2 }  \Bigg].
\end{align}
Here, we call the one in Eq.~\eqref{eq:First}  (Eq.~\eqref{eq:Second}) the first (second) term and note that $s^2(\Omega) = s_1^2(\Omega) +s_2^2(\Omega) $. In the limit of vanishing absorption, the first term goes to zero whereas the second one becomes equal to the resonant term in Eq.~\eqref{eq33:S2PPFinal}. Hence, one can say that the contributions stemming from the first term represent a novel contribution to the signal stemming from absorption effects. \\
 These two different terms are plotted together with the experimental data and the full theoretical result (Eq.~\eqref{eq33:sOmegaAbs} ) in Fig.~2. We see that most contributions accessed in the experiment reported in Ref.~\cite{Benea-Chelmus2019} stem from the first term. This shows the crucial importance of taking absorption effects into account. 

\begin{figure}
\centering
\includegraphics[width=0.5\textwidth]{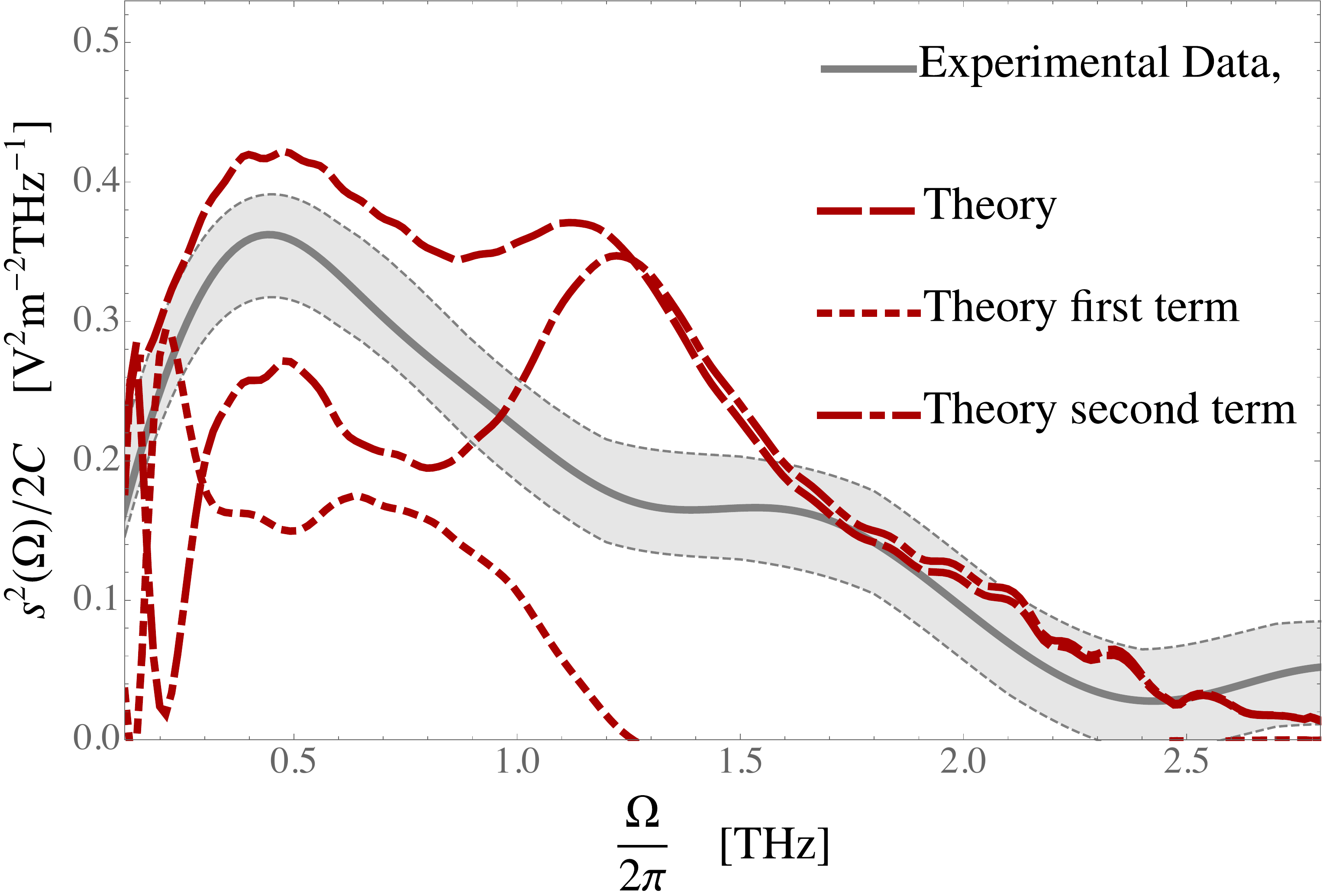}
\label{fig:Origin}
\caption{ \textit{Comparison of experimental data to the different terms in Eq.~\eqref{eq33:sOmegaAbs}:} As in the bottom plot of Fig.~3, we show the experimental data taken from Ref.~\cite{Benea-Chelmus2019} and compare it to our theoretical result in Eq.~\eqref{eq33:sOmegaAbs}. We further individually plot the first (Eq.~\eqref{eq:First}) and second (Eq.~\eqref{eq:Second}) term of the theoretical result in order to show which one contributes most to the measured signal.   }
\end{figure}

\end{widetext}

\end{document}